Prediction of Silicon-Based Layered Structures for Optoelectronic Applications

Wei Luo,[1] Yanming Ma,[2] Xingao Gong,[1] and Hongjun Xiang[1]*

[1]Key Laboratory of Computational Physical Sciences (Ministry of Education), State Key Laboratory of Surface Physics, and Department of Physics, Fudan University, Shanghai 200433, P. R. China

[2]State Key Lab of Superhard Materials, Jilin University, Changchun 130012, China

Email: hxiang@fudan.edu.cn

**ABSTRACT**

A method based on the particle swarm optimization (PSO) algorithm is presented to design quasi-two-dimensional (Q2D) materials. With this development, various single-layer and bi-layer materials in C, Si, Ge, Sn, and Pb were predicted. A new Si bi-layer structure is found to have a much-favored energy than the previously widely accepted configuration. Both single-layer and bi-layer Si materials have small band gaps, limiting their usages in optoelectronic applications. Hydrogenation has therefore been used to tune the electronic and optical properties of Si layers. We discover two hydrogenated materials of layered $Si_8H_2$ and $Si_6H_2$ possessing quasi-direct band gaps of 0.75 eV and 1.59 eV, respectively. Their potential applications for light emitting diode and photovoltaics are proposed and discussed. Our study opened up the possibility of hydrogenated Si layered materials as next-generation optoelectronic devices.

**INTRODUCTION**

In recent years, two-dimensional (2D) or Q2D materials have attracted numerous interests for their fascinating properties. Graphene, a single-layer of carbon atoms with honeycomb configuration, has been widely studied[1-4] due to its novel Dirac-like electronic properties. It has rapidly become a candidate for the next generation of faster and smaller electronic devices. Besides graphene, other layered materials[5,6,7] with excellent properties were also discovered. Monolayer $MoS_2$ with a direct band gap[8,9] can be used to construct interband tunnel field-effect transistors (FET),[10] which have lower power consumption than classical transistors. Recently, few-layer black phosphorus crystals with thickness down to a few nanometers have been fabricated. Researchers show that it is also a good material for fabricating field-effect transistors (FET).[5] Coleman *et al.*[11] reported that a variety of inorganic layered materials can be obtained through a straightforward liquid exfoliation technique.

Due to the high stability, high abundance, and the existence of an excellent compatible oxide ($SiO_2$), Si is the leading material of microelectronic devices. Because of the excellent compatibility with the mature Si based microelectronics industry and high abundance of Si, Si-based layered materials may be the most promising layered materials for realistic applications. Although the majority of solar cells fabricated to date have been based on three-dimensional (3D) bulk diamond Si,[12] it is well known that 3D bulk Si is not an ideal material for optoelectronic applications because Si is an indirect band gap (1.1 eV) semiconductor and there is a large energy difference between the direct gap (3.4 eV) and the indirect gap.[13] Thus, the discovery of Si-based layered materials with excellent optoelectronic properties might lead to revolution in future optoelectronics technology.

Like graphene, single-layer Si has a honeycomb structure with the linear dispersion of the band structure near the Dirac point.[14-19] Moreover, it was identified as a topological insulator[20,21] (TI) with a tiny band gap (1.55 meV), which is too small for optoelectronic applications. In the same time, singer-layer silicene itself is unstable. It must be grown on metallic substates[22,23] and can not be exfoliated from substrates. Furthermore, substrates change the electronic properties of silicene.[24-26] Thus, it is highly desirable to discover new free-standing stable Si-based layered materials with exceptional optoelectronic properties.

In our work, we propose a new general global optimization method to predict quasi-2D structures based on the PSO technique as implemented in the Crystal structure Analysis by Particle Swarm Optimization (CALYPSO) code.[27,28] With this development, a new Si bi-layer structure with lower energy than the widely accepted lowest energy configuration has been predicted. Furthermore, we find two layered structures, namely $Si_8H_2$-Pm11 and $Si_6H_2$-Pmm2. They have quasi-direct gaps of 0.75 eV and 1.59 eV, respectively. These new structures are promising candidates for optoelectronic applications.

**COMPUTATIONAL METHODS**

**PSO Algorithm for Quasi-2D System.** PSO is designed to solve problems related to multidimensional optimization,[29] which is inspired by the social behavior of birds flocking or fish schooling. Recently, the PSO algorithm was adopted to successfully predict 3D crystal structure as implemented in CALYPSO code.[27,28] Previously, we proposed a modified PSO algorithm for predicting single-atom thick planar 2D crystals, such as graphene and BN sheet.[7] This approach was subsequently extended to search multi-layer structures.[30] In this later development, it requires the number of layers and number of atoms in each layer as the input[30] when searching a multi-layer Q2D structure. However, this information is not always known prior to the search of a new Q2D system, which limits its usage. Here, we propose a more general way to predict Q2D systems based on the PSO algorithm.[27] With this development, prediction of Q2D materials without specifying the number of layers and the chemical composition per layer becomes feasible.

In our implementation, we generate random Q2D structures whose thickness along the c-axis is smaller than a given thickness $d$. We note that specifying the thickness $d$ in the structure searching is more general and unambiguous than specifying the number of layers. A Q2D crystal can be characterized by 80 layer space groups, different from the planar 2D crystals characterized by 17 planar space groups. To generate a random Q2D structure, we first randomly select a layer group. The lateral lattice parameters and atomic positions are then randomly generated but confined within the chosen layer group symmetry. The generation of random structures ensures unbiased sampling of the energy landscape. The explicit application of symmetric constraints leads to significantly reduced search space and optimization variables, and thus fastens global structural convergence. Subsequently, local optimization including the atomic coordinates and lateral lattice parameters is performed for each of the initial structures. In the next generation, a certain number of new structures (the best 60% of the population size) are generated by PSO. The other structures are generated randomly, which is critical to increase the structure diversity. Within the PSO scheme,[27] a structure in the searching phase space is regarded as a particle. A set of particles (structures) is called a population or a generation. The positions of the particle are updated according to the following equation:

$$x_{i,j}^{t+1} = x_{i,j}^t + v_{i,j}^{t+1}$$

Where x and v are the position and velocity, respectively [i is the atom index, j refers to the dimension of structure ($1 \leq j \leq 3$), and it is the generation index]. The new velocity of each particle is calculated on the basis of its previous location $x_{i,j}^t$ before optimization, previous velocity $v_{i,j}^t$, current location $pbest_{i,j}^t$ with an achieved best fitness (i.e., lowest energy), and the population global location $gbest_{i,j}^t$ with the best fitness value for the entire population:

$$v_{i,j}^{t+1} = wv_{i,j}^t + c_1 r_1 (pbest_{i,j}^t - x_{i,j}^t) + c_2 r_2 (gbest_{i,j}^t - x_{i,j}^t)$$

Where ω (in the range of 0.9-0.4) denotes the inertia weight, $c_1 = 2$ and $c_2 = 2$, $r_1$ and $r_2$ are two separately generated random numbers and uniformly distributed in the range [0, 1]. The initial velocity is generated randomly. All the structures produced by the PSO operation are then relaxed to the local minimum. When we obtain the new quasi-2D structure by the PSO operation or random generation, we make sure that the thickness of the quasi-2D structure is smaller than the given thickness *d*. Usually, tens of iterations are simulated to make sure that the lowest-energy structures are found.

To test the efficiency of this new method, we search some known Q2D systems.

We set the population size to 30 and the number of generations to 20. Without giving any prior structure information except for the whole chemical composition, we find the known structures for layered $MoS_2$, GaS, and $Bi_2Se_3$ within 3, 2, and 11 generations, respectively. These tests indicate that our approach and its implementation are efficient.

In our simulation for the silicon-based Q2D systems, the total number of atoms is fixed to be no more than 12 in the unit cell. We only consider even number atoms since many tests show that systems with odd number atoms usually have higher total energy. We consider five different thickness $d$ (usually between 1 to 5 Å) for each system. In addition, we repeat twice of each calculation in order to make results reliable.

**DFT Calculations.** In this work, density functional theory (DFT) method is used for structural relaxation and electronic structure calculation. The ion-electron interaction is treated by the projector augmented-wave (PAW[31]) technique which performed in the Vienna ab initio simulation package (VASP[32]). In the DFT plane-wave calculations, we use the local density approximation (LDA[33,34]). The 2D k-mesh is generated by the Monkhorst-Pack scheme which depends on lattice constant. For relaxed structures, the atomic forces are less than 0.01 eV/Å. To avoid interactions between different quasi-ab-planes, we set the vacuum thickness to 12 angstrom. As we know, LDA underestimate band gap, so we adopt the HSE06 functional to calculate the electronic and optical properties.[35] To make sure the dynamical stability of the obtained structures, we use the finite displacement method[36] as implemented in the PHONOPY code[37] to calculate the phonon frequencies.

**Results and Discussion**

We predict the single-layer and bi-layer structures of the silicon related group IV elements systematically. Our results are listed in TABLE I.

**Single-Layer and Bi-Layer Si.** If the thickness is less than the radius[38] of the element, we call it single-layer structure. For single-layer Si, our results agree with previous report[20]. It is the honeycomb silicene structure with a buckling of about 0.442 Å. Moreover, silicene is a TI in which Quantum Spin Hall Effect (QSHE) may be realized.[20] For bi-layer Si, it is widely accepted the AA stacking configuration [shown in figure 1(b), referred as AA-Si) is the most stable structure.[39] Recently, Florian Gimbert *et al.*[40] reported that Q2D Si may form the $MoS_2$–type structure. However, its total energy seems too high to be synthesized experimentally. From our PSO simulation, we find a new bi-layer structure of Si with the layer group Cmme, namely Si-Cmme. The LDA total energy of Si-Cmme is lower than AA-Si by 41 meV/atom. We also use the Perdew-Burke-Ernzerhof (PBE[41]) exchange-correlation function to calculate total energy and find similar results. From Figs. 1(b) and 1(a), we can easily see that AA-Si has the $C_3$ rotational symmetry, while the $C_3$ symmetry is broken in the Si-Cmme. Neither of these two structures has buckling. The simulated occupied STM images [Figs. 1(c) and 1(d)] of Si-Cmme and AA-Si indicate that our prediction of the Si-Cmme structure may be easily confirmed by the STM experiment.

We calculate the electronic structures of AA-Si and Si-Cmme with the HSE06 functional. Band structures are shown in figure 4(a) and figure 4(b). From figure 4(a), we find that AA-Si is an indirect semiconductor with a band gap of about 130 meV, in agreement with previous result.[39] However, Si-Cmme is metallic with large dispersions of the valence band maximum (VBM) and conduction band minimum (CBM) states. Taking the vacuum level as the reference, we find that the CBM of Si-Cmme is much lower (by almost 1.0 eV) than that of the AA-Si. This can be seen from the band edge positions when the AA-Si structure evolves into the Si-Cmme structure (see Fig. S4 of the supporting information). It should be noted that the CBM of Si-Cmme is below the Fermi level, thus is occupied. This is in accord with the lower energy of the Si-Cmme phase. In addition, calculations show that Si-Cmme is dynamically and thermally stable (see supporting information). Since each Si atom in the bi-layer Si-Cmme is four-fold coordinated, Si-Cmme is much more stable (by 270 meV/atom) than the single-layer silicene.

**Single-Layer and Bi-Layer of other group elements.** For carbon, our results agree with previous study[1] very well. So we do not list results of C in TABLE I. Like silicene, single-layer Ge and Sn have a honeycomb configuration and are TIs. The band gaps are larger than silicene, because the spin-orbit coupling is stronger than that in Si. For Pb, we find that there is no dynamically stable single-layer structure. Similar to bi-layer Si, we find bi-layer Ge also has the Cmme configuration which has a lower energy by 21 meV/atom than the AA-Ge bi-layer structure. For bi-layer Sn and bi-layer Pb, the lowest structure belongs to the layer group $P\bar{3}m1$ (see figure 2), which is different from the cases of Si and Ge. All these bi-layer structures are found to be metallic.

Our above simulations show that both single-layer and bilayer Si structures are not suitable for the optical and electronic applications because both structures do not have an appropriate band gap. It was shown that chemical functionalization can be used to tune or modify the electronic and optical properties of 2D and quasi-2D systems.[39,42-45] In order to identify Q2D Si-based systems for optoelectronic applications, we perform global search of hydrogenated Q2D Si structures. We consider four different compositions, i.e., $Si_8H_2$, $Si_6H_2$, $Si_4H_2$, and $Si_4H_4$.

**Hydrogenate Si Layered Structure ($Si_8H_2$-Pm11).** For $Si_8H_2$, we find that the lowest energy structure among all structures with thickness less than 6.0 Å is $Si_8H_2$-Pm11, as shown in figure 3(a). Similar to the proposed $Si_{24}H_6$ structure[39], the hydrogen atoms only absorb at the bottom side in $Si_8H_2$-Pm11. However, $Si_8H_2$-Pm11 has a lower energy by 73 meV/atom than the previous proposed structure.[39] Except for the two three-fold coordinated Si atoms on the top surface, all the other Si atoms are four-fold coordinated. The two three-fold coordinated Si atoms form a one-dimensional zigzag chain. As can be seen from the band structure shown in figure 4(c), the direct band gap and indirect band gap are 0.75 eV and 0.70 eV, respectively, which is close to the optimal wavelength (1.55 μm ≈ 0.8 eV) for the optical fiber communications. From the band decomposed charge density [figure 5(a)], we find that the VBM and CBM states are mainly contributed by the zigzag chain formed by the three-fold coordinated Si atoms. Furthermore, we compute the imaginary part $\varepsilon_2$ of the dielectric function to find that the direct gap transition is dipole-allowed. This is expected since $Si_8H_2$-Pm11 belongs to the Cm point group and the transition between any two eigenstates of a Cm system is dipole-allowed. The optical transition near the band edge is found to be rather strong. This is because both VBM and CBM states are 1D-like, resulting in the Van Hove-like singularity in the density of state near the band edge. Our calculations show that $Si_8H_2$-Pm11 is dynamically and thermally stable (see supporting information).

**Hydrogenate Si Layered Structure ($Si_6H_2$-Pmm2).** We find that the structure (i.e., $Si_6H_2$-Pmm2) shown in figure 3(b) has the lowest energy among all Q2D structures with the thickness less than 6.1 Å. In $Si_6H_2$-Pmm2, there are four different kinds of Si atoms. Each Si atom is four-fold coordinated, indicating that this structure is highly stable. For type-I, type-II and type-IV Si atoms, all the four nearest neighbors are Si atoms. The type-III Si atom forms a covalent bond with one H atom. The computed HSE06 band structure [figure 4(d)] suggests that $Si_6H_2$-Pmm2 also has a quasi-direct band gap. The direct band gap at $\Gamma$ and indirect band gap are 1.59 eV and 1.52 eV, respectively, which is near the optimal value (about 1.5 eV) for solar-energy absorber applications. It is interesting to note that there is a Si triangle in Q2D $Si_6H_2$-Pmm2, similar to the predicted 3D structure $Si_{20}$[46] with a quasi-direct band gap near 1.5 eV. It appears that the presence of Si triangles is relevant to the desirable band gap value. The band decomposed charge density plot [figure 5(b)] shows that the band edge states, in particular the CBM state, distribute mostly around the Si triangle. The computed imaginary part $\varepsilon_2$ of the dielectric function shows that direct-gap transition in $Si_6H_2$-Pmm2 is dipole allowed (see figure 6). This can be easily understood by the group theory analysis. The point group of $Si_6H_2$-Pmm2 is $C_{2v}$ with an in-plane two-fold rotational axis. Detailed analysis shows that both the VBM and CBM states at Γ belong to the $A_1$ irreducible representation. Thus, the transition between the VBM and CBM states can be induced by a light with the electric field along the two-fold rotational axis of the system. The good optical absorption and optimal direct gap suggest that $Si_6H_2$-Pmm2 is a good nanoscale solar cell absorber. The computed phonon frequencies (see supporting information) indicate that $Si_6H_2$-Pmm2 is dynamically stable. Furthermore, molecular dynamics (MD) simulation indicate it is thermally stable at least up to 800 K (see supporting information).

**Hydrogenate Si Layered Structure ($Si_4H_2$-$P\bar{3}m1$ and $Si_4H_4$-Pman).** For $Si_4H_2$, we find the lowest energy layered structure is $Si_4H_2$-$P\bar{3}m1$, as shown in supporting information. We can see each Si atom is four-fold coordinated, forming the good $sp^3$ hybridization, which contribute to the large indirect band gap (about 2.1 eV) and low total energy. For $Si_4H_4$, it is widely accepted that the $Si_4H_4$ structure based on the silicene structure ($Si_4H_4$-$P\bar{3}m1$) is the most stable one. Here we find a new structure ($Si_4H_4$-Pman, see supporting information) based on the black phosphorus structure has a lower energy by about 4 meV/atom than that based on the silicene structure (see supporting information).

From our simulations on hydrogenated Si layered structures, we find that the band gap tends to increase with the concentration of hydrogen because the Si atoms can form better $sp^3$ hybridization when there are more H atoms. Furthermore, we find that hydrogenation improves the thermal stability of the Si layered structures. For silicene and Si-Cmme, the melting points are about 500 K and 600 K (see supporting information). While for $Si_8H_2$-Pm11 and $Si_6H_2$-Pmm2, the melting points are about 800 K and 900 K.

**Conclusion**

In summary, we have developed a new method based on the PSO algorithm to predict the Q2D crystal structure. Combining this method with first principle calculations, we predict systematically the single-layer and bi-layer structures of C, Si, Ge, Sn and Pb. A new stack configuration of bi-layer Si is revealed to have a lower energy than the previous known structures. Furthermore, we systematically search the hydrogenated Si layered structures, i.e., $Si_8H_2$, $Si_4H_2$, $Si_6H_2$, and $Si_4H_4$. In particular, we find that $Si_8H_2$-Pm11 and $Si_6H_2$-Pmm2 are stable and have desirable optical properties, thus are excellent candidates for generating light with optimal wavelength for the optical fiber communications and absorbing light in a solar cell, respectively.

**Supporting Information.** Phonon dispersions of Si-Cmme, $Si_8H_2$-Pm11, $Si_6H_2$-Pmm2; thermal stability of silicene, Si-Cmme, $Si_8H_2$-Pm11, $Si_6H_2$-Pmm2; lowest energy layered structure of $Si_4H_2$ and $Si_4H_4$; evolution of the CBM, VBM, and total energy from the AA-Si structure to the Si-Cmme structure; geometrical coordinates and energies for optimized structures; Complete reference 11. This information is available free of charge via the Internet at http://pubs.acs.org.

**Acknowledgments**

Work was supported by NSFC, the Special Funds for Major State Basic Research, FANEDD, NCET-10-0351, Research Program of Shanghai Municipality and MOE, Program for Professor of Special Appointment (Eastern Scholar), and Fok Ying Tung Education Foundation.

TABLE I: Single-layer and bi-layer structures of Si, Ge, Sn, and Pb predicted from our global structure optimizations. The total energy is calculated by using LDA. Some configurations (i.e., Cmme, $P\bar{3}m1$) are shown in Fig. 1(a) and Fig. 2. There are no dynamically stable single-layer Pb structures.

| Element | Layer number | Energy(eV) | Thickness(Å) | Configuration | Property |
|---|---|---|---|---|---|
| Si | 1 | -5.2265 | 0.442 | Silicene | TI |
| Si | 2 | -5.4960 | 2.249 | Cmme-Si | Metal |
| Ge | 1 | -4.5215 | 0.624 | Silicene Like | TI |
| Ge | 2 | -4.7281 | 2.405 | Cmme-Ge | Metal |
| Sn | 1 | -3.8454 | 0.806 | Silicene Like | TI |
| Sn | 2 | -4.1785 | 2.600 | $P\bar{3}m1$-Sn | Metal |
| Pb | 1 | Not stable | | | |
| Pb | 2 | -4.0523 | 2.603 | $P\bar{3}m1$-Pb | Metal |

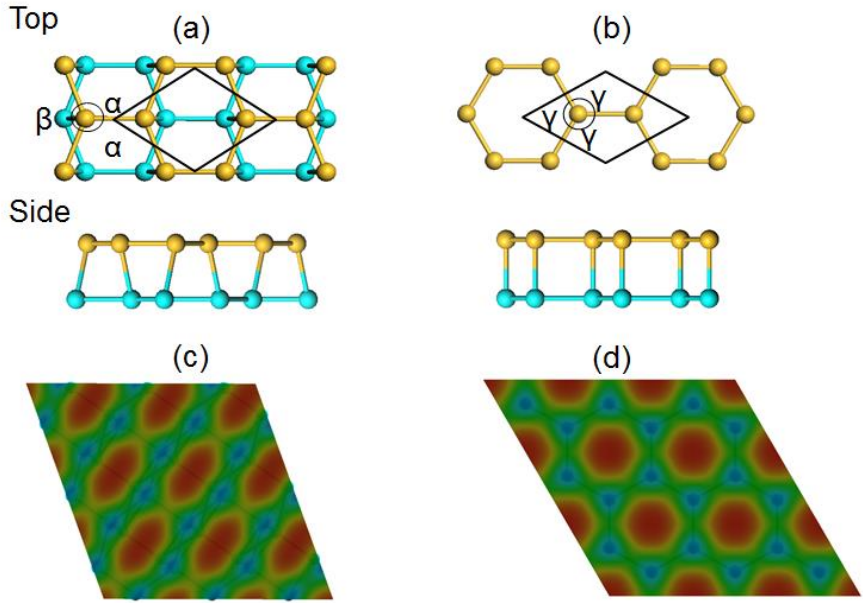

FIGURE 1: (a) Top and side views of Si-Cmme. (b) Top and side views of AA-Si. (c) The simulated occupied STM image of Si-Cmme. (d) The simulated occupied STM image of AA-Si. α=110.2°, β=139.6°, γ=120°. The Si atoms of the top layer and those of the bottom layer are indicated by different colors. The STM images are obtained by summing up all the local density of states from -1.0 eV to the Fermi level. From figure (b), we can easily see that AA-Si has $C_3$ symmetry, different from the case of Si-Cmme. The lateral unit cell is denoted by the black lines.

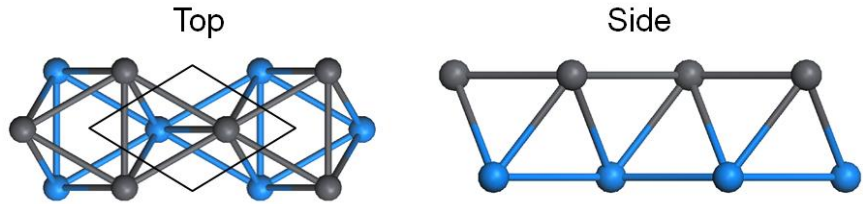

FIGURE 2: The top and side views of bi-layer $P\bar{3}m1$-Sn ($P\bar{3}m1$-Pb is the same as $P\bar{3}m1$-Sn). The bilayer structure is formed by stacking two triangle layers.

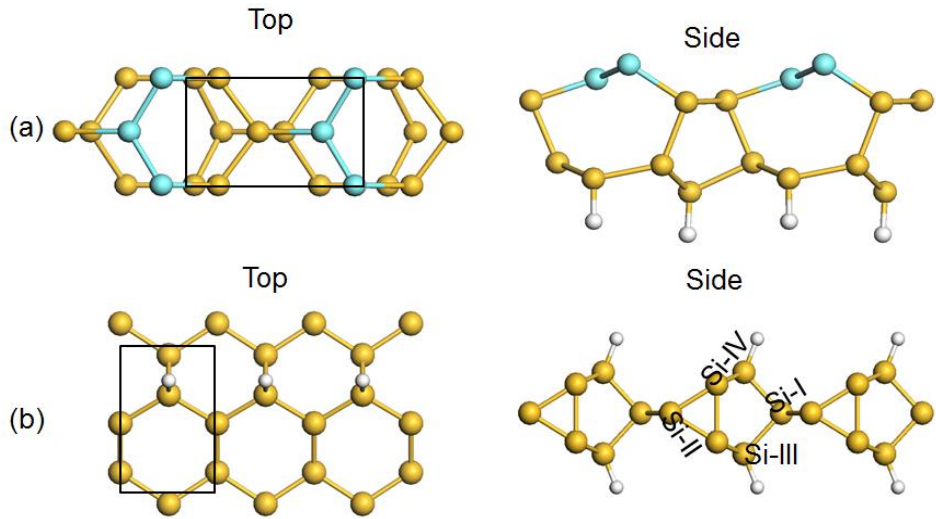

FIGURE 3: (a) The top and side views of $Si_8H_2$-Pm11. Three-fold coordinated and four-fold coordinated Si atoms are marked by aquamarine and yellow respectively. These three-fold coordinated Si atoms form a one-dimensional zigzag chain. (b) The top and side views of $Si_6H_2$-Pmm2. Each Si atom is four-fold coordinated. For type-I, type-II, and type-IV Si atoms, all the four nearest neighbors are Si atoms. The type-III Si atom forms a covalent bond with one H atom. Two type-IV Si atoms and one

type-II Si atom form a triangle.

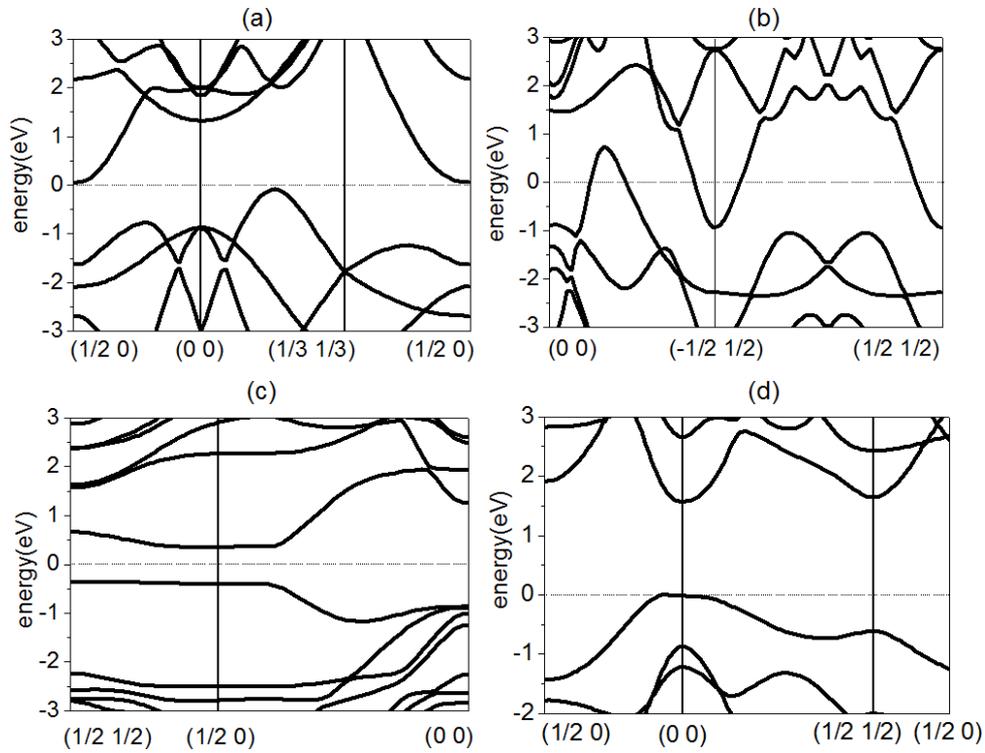

FIGURE 4: (a) Band structure of AA-Si from the HSE calculations. We can see that it is an indirect semiconductor with a very small band gap. (b) Band structure of Si-Cmme. Unlike AA-Si, it is a semi-metal. (c) Band structure of $Si_8H_2$-Pm11. The VBM and CBM are almost flat along the direction that is perpendicular to the 1D zigzag chains formed by three-fold coordinated Si atoms. (d) Band structure of $Si_6H_2$-Pmm2. It is a semiconductor with quasi-direct band gap 1.59 eV.

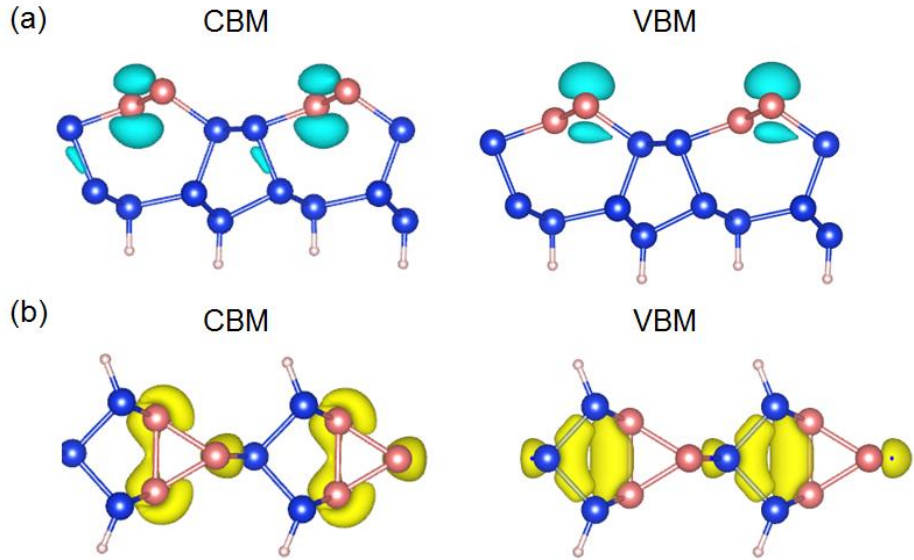

FIGURE 5: (a) The partial charge density of $Si_8H_2$-Pm11. Red atoms are three-fold coordinate Si atoms. From this figure, we can see that the CBM and VBM states are mainly contributed by three-fold coordinate Si atoms. (b) The partial charge density of $Si_6H_2$-Pmm2. Three four-fold coordinated red Si atoms form a triangle. Similar to $Si_{20}$, the band edge states distribute mainly around the Si triangle.

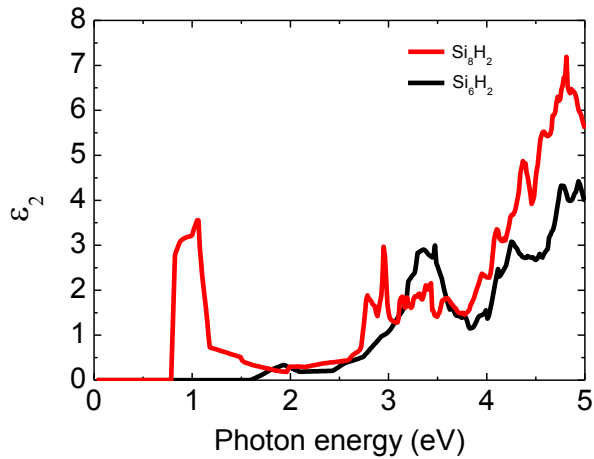

FIGURE 6: Imaginary part of dielectric functions from the HSE06 calculations for $Si_6H_2$-Pmm2 and $Si_8H_2$-Pm11.

TOC graphic

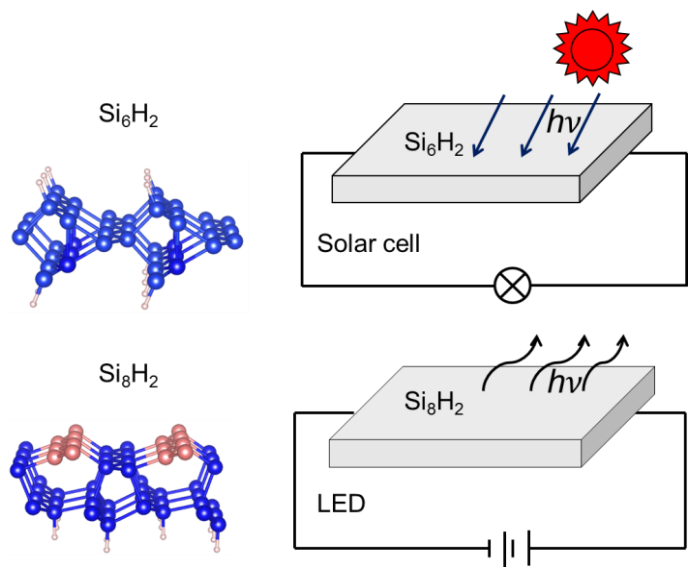